\documentclass[aps,prd,reprint,amsmath,amssymb,superscriptaddress,floatfix,nofootinbib,showkeys,showpacs,preprintnumbers]{revtex4-2} 

\usepackage[dvipsnames]{xcolor}
\usepackage{hyperref}
\hypersetup{    
  colorlinks      = true,
  linkcolor       = {blue},
  linkbordercolor = {white},
  citecolor       = {blue},
  citebordercolor = {white},
  urlcolor        = {blue},
  urlbordercolor  = {white},
}

\usepackage{amsmath, amssymb} 
\usepackage{xspace} 
\usepackage{mathtools, tabu}
\usepackage{newtxtext, newtxmath}
\usepackage[T1]{fontenc}
\usepackage{graphicx} 
\usepackage{orcidlink}

\begin{document}

\newcommand{\think}[1]{\textcolor{NavyBlue}{#1}}
\newcommand{\todo}[1]{\textcolor{BurntOrange}{#1}}
\newcommand{\newtext}[1]{\textcolor{ForestGreen}{#1}}
\newcommand{\titus}[1]{\textcolor{Red}{#1}}


\newcommand{\redmapper}{redMaPPer\xspace}

\newcommand{\avg}[1]{\left\langle{#1}\right\rangle}
\newcommand{\beq}{\begin{equation}}
\newcommand{\eeq}{\end{equation}}

\newcommand{\OmegaM}{\Omega_\mathrm{m}}

\newcommand{\rp}{r_{\rm p}}
\newcommand{\rhom}{\rho_{\rm m}}
\newcommand{\xihm}{\xi_{\rm hm}}
\newcommand{\rs}{r_{\rm s}}
\newcommand{\pimax}{\pi_{\rm max}}

\newcommand{\hiMpc}{h^{-1} \rm cMpc}
\newcommand{\hikpc}{h^{-1} \rm kpc}
\newcommand{\hiMsun}{h^{-1} M_\odot}
\newcommand{\Msun}{M_\odot}

\newcommand{\Ncen}{N_{\rm cen}}
\newcommand{\Nsat}{N_{\rm sat}}

\title{Impact of projection-induced optical selection bias on the weak lensing mass calibration of galaxy~clusters}

\author{Titus Nyarko Nde\textsuperscript{\orcidlink{0000-0002-0089-0774}}}
\altaffiliation[Corresponding  Author: ]{\href{mailto:titusnyarkonde@u.boisestate.edu}{titusnyarkonde@u.boisestate.edu}}
\affiliation{School of Computing, Boise State University, Boise, ID, 83725, USA}
\author{Hao-Yi Wu}
\affiliation{Department of Physics, Southern Methodist University, Dallas, TX 75205, USA}
\author{Shulei Cao}
\affiliation{Department of Physics, Southern Methodist University, Dallas, TX 75205, USA}
\author{Gladys~Muthoni~Kamau}
\affiliation{School of Computing, Boise State University, Boise, ID, 83725, USA}
\author{Andrius Tamosiunas}
\affiliation{Instituto de F\'{\i}sica Te\'orica (IFT), UAM--CSIC, Campus de Cantoblanco UAM, Madrid, Spain}
\affiliation{CERCA/ISO, Department of Physics, Case Western Reserve University, Cleveland, Ohio 44106, USA}
\author{Chun-Hao To}
\affiliation{Department of Astronomy and Astrophysics, University of Chicago, Chicago, IL 60637, USA}
\author{Conghao Zhou}
\affiliation{Physics Department, University of California, Santa Cruz, CA 95064, USA\\
Santa Cruz Institute for Particle Physics, Santa Cruz, CA 95064, USA}

\date{\today}

\begin{abstract}
 
Weak gravitational lensing signals of optically identified clusters are impacted by a selection bias---halo triaxiality and large-scale structure along the line of sight simultaneously boost the lensing signal and richness (the inferred number of galaxies associated with a cluster).  As a result, a cluster sample selected by richness has a mean lensing signal higher than expected from its mean mass, and the inferred mass will be biased high.  This selection bias is currently limiting the accuracy of cosmological parameters derived from optical clusters.  In this paper, we quantify the bias in mass calibration due to this selection bias.  Using two simulations, MiniUchuu and Cardinal, with different galaxy models and cluster finders, we find that the selection bias leads to an overestimation of lensing mass at the $20-50\%$ level, with a larger bias ($20-80\%$) for large-scale lensing ($>3$ Mpc).  Even with a moderate projection model, this selection bias significantly outweighs other currently known cluster lensing systematics. This work confirms the need to account for this bias in future optical cluster cosmology analyses, and we discuss strategies for mitigating this bias. 

\end{abstract}
\maketitle

\section{Introduction}

The number counts of galaxy clusters as a function of mass across cosmic time are sensitive to the mean matter density $\OmegaM$, the density fluctuation parameter $\sigma_8$, and dark energy \cite{Haiman01, Holder01, LimaHu04, Voit05, Allen2011, Weinberg13}.  Wide-field optical imaging surveys \cite{DES2005, Euclid, LSST_DESC2012, Akeson19, Eifler21HLS} simultaneously identify clusters by their galaxy content and provide weak gravitational lensing signals for mass calibration, or equivalently, the calibration of the observable--mass relation \cite{Rozo2010, Costanzi19SDSS, To2021, DESY1CL, Lesci22, Sunayama23}.  These surveys can identify clusters down to low masses, providing statistical power comparable to that of cosmic shear \cite{OguriTakada11, YooSeljak12, Weinberg13, Wu21}.  In cluster cosmology analyses, the accuracy of cosmological constraints depends on the accuracy of the mass calibration \cite{Simet2017, Melchior2017, McClintock2019, Murata18, Pratt19}.

Currently, the accuracy of mass calibration for optically selected clusters is limited by a selection bias caused by line-of-sight projection effects \cite{Angulo2012, Busch2017, Sunayama2020, Wu2022}, as described below.  In optical surveys, we assign each cluster a {\em richness} value $\lambda$, defined as the membership-weighted number of galaxies associated with a cluster. In the presence of distance uncertainties, galaxies physically unassociated with the cluster but along the line of sight can be mistaken as cluster members and boost cluster richness \cite{Costanzi2019, Myles2021, Lee2025, Cao2025, Myles25projection}.  At the same time, the lensing signal of clusters is determined by the projected matter density \cite{Becker2011, Bahe12, Umetsu2020}. As a result, if a cluster happens to have an excess of galaxies and matter along the line of sight (due to halo triaxiality or filaments), its richness and lensing signals will be coherently boosted. In addition, since low-mass halos are much more abundant than high-mass halos, a cluster sample defined by a richness range has more low-mass halos with up-scattered richness than high-mass halos with down-scattered richness \cite{LimaHu05}.  Given that up-scattered richness is related to up-scattered lensing, the mean lensing signal for clusters in a richness bin is biased high compared to expectation based on the cluster masses \cite{Sunayama2020, Wu2022, Zhang23triaxiality}.

In this paper, we use simulated cluster catalogs to quantify the impact of this projection-induced selection bias on weak lensing mass calibration.  We use two mock catalogs with different galaxy--halo connection models and cluster finders.  The MiniUchuu-based mock uses a halo occupation distribution (HOD) model \cite{BerlindWeinberg02, Cooray2002, Zheng05, Zehavi11} and a counts-in-cylinder cluster finding algorithm \cite{Sunayama2020, Zeng2023, Salcedo2024, Lee2025}, while the Cardinal mock \cite{To2024} is based on the ADDGALS (Adding Density Dependent GAlaxies to Lightcone Simulations) galaxy model \cite{Wechsler2022} and the \redmapper (red-sequence Matched-filter Probabilistic Percolation) cluster finder \cite{Rykoff2014, Rykoff2016}. We simulate the stacked lensing signal and perform a mock mass calibration.  Our results indicate that the selection bias leads to $20-50\%$ bias in mass calibration, making it the dominant systematic uncertainty in optical cluster cosmology.

This work is built on several recent studies on projection effects \cite{Osato18, Costanzi2019, Sunayama2020, Myles2021, Wu2022, Zhang23triaxiality, Myles25projection, Lee2025, Cao2025}.  Refs.~\cite{Costanzi2019, Cao2025} quantify the impact of correlated and uncorrelated line-of-sight structure on cluster richness.  Refs.~\cite{Sunayama2020, Wu2022} use $N$-body simulations populated with galaxies to quantify the impact of projection effects on stacked lensing.  Refs.~\cite{Myles2021, Myles25projection} use spectroscopic redshifts of member galaxies of redMaPPer clusters to quantify the fraction of projected members.

This paper is organized as follows. We describe the basic formalism of cluster lensing in Sec.~\ref{sec:basics} and present the two cluster simulations in Sec.~\ref{sec:sims}.  Section~\ref{sec:mass_bias} presents our results of weak lensing mass calibration. We summarize and discuss our results in Sec.~\ref{sec:summary}.  Throughout this paper, Mpc refers to physical Mpc (used in lensing measurements and fitting), while cMpc refers to comoving Mpc (used in $N$-body simulations).

\section{Basic formalism for cluster lensing}\label{sec:basics}

In this section, we describe the basic formalism for cluster lensing, following \cite{McClintock2019} and the {\tt cluster$\_$toolkit} software\footnote{\href{https://cluster-toolkit.readthedocs.io/}{https://cluster-toolkit.readthedocs.io/}}.  We use the halo--matter correlation function $\xihm$ to model the mass distribution around clusters.  
The 1-halo term is described by the Navarro--Frenk--White \cite{Navarro1997} profile 
\begin{align}
\xi_\mathrm{1h}(r, M) = \frac{\rho_\mathrm{NFW}(r, M)}{\rho_{\rm m}} - 1
= \frac{\delta_{\rm m}}{(r/\rs)\left(1 + r/\rs\right)^2}  - 1,
\end{align}
where $r$ is the 3D distance to the halo center, and $\rho_{\rm m}$ is the mean matter density of the Universe.  Here $\rs$ is the scale radius, which is related to halo concentration via $c_{\Delta} = R_{\Delta}/\rs$ for a given spherical overdensity mass definition ($\Delta=$ 200m or vir); $\delta_{\rm m}$ is the characteristic overdensity set by the concentration and the total mass of the halo \cite{Wright1999}.  

The 2-halo term is modeled by the matter correlation function and a linear halo bias
\begin{align}
    \xi_{\mathrm{2h}}(r, M) = b(M)\xi_{\rm mm}(r),
\end{align}
where $b(M)$ is the halo bias calculated with the fitting formula in \cite{Tinker2010}.  The matter correlation function $\xi_{\rm mm}$ is calculated with the {\tt halofit} \cite{Takahashi12halofit} non-linear matter power spectrum from {\tt class} \cite{Blas2011}.  For the 1-halo and 2-halo transition, we follow \cite{Zu2014},
\begin{align} 
    \xi_\mathrm{hm} = 
    {\rm max}\big(
        \xi_{\mathrm{1h}},~
        \xi_{\mathrm{2h}}
        \big) .
\end{align}

For a given halo--matter correlation function, we compute the surface density profile
\begin{align}
    \Sigma(\rp) = \rho_{\rm m}\int_{-\infty}^{+\infty} \xihm\left(r = \sqrt{\rp^2 + \pi^2}\right){\rm d}\pi,
\end{align}
where $\rp$ is the 2D projected distance to the halo center, and $\pi$ is the distance along the line of sight.

The weak lensing signal is given by the excess surface mass density profile 
\begin{align}
    \label{eq:DeltaSigma}
    \Delta\Sigma(\rp) = \overline{\Sigma}(<\rp) - \Sigma(\rp),
\end{align}
where the first term on the right is the mean surface density within $\rp$,
\begin{align}
    \overline{\Sigma}(<\rp)
    = \frac{2}{\rp^2}\int_0^{\rp} \rp^{\prime} \Sigma(\rp^{\prime}){\rm d}\rp^{\prime} .
\end{align}
To simulate the lensing signal of source galaxies in radial bins, we use the radially averaged profiles 
\begin{align}
    \label{eq:DeltaSigma}
    \overline{\Delta\Sigma}= \frac{2}{r_{\rm p,1}^2-r_{\rm p,2}^2}\int^{r_{\rm p,2}}_{r_{\rm p,1}}\rp^{\prime}\Delta\Sigma(\rp^{\prime}){\rm d}\rp^{\prime}.
\end{align}
We use comoving units when calculating correlation functions and convert the resulting lensing signal to physical units.

\begin{figure*}
\centering
\includegraphics[width=2\columnwidth]{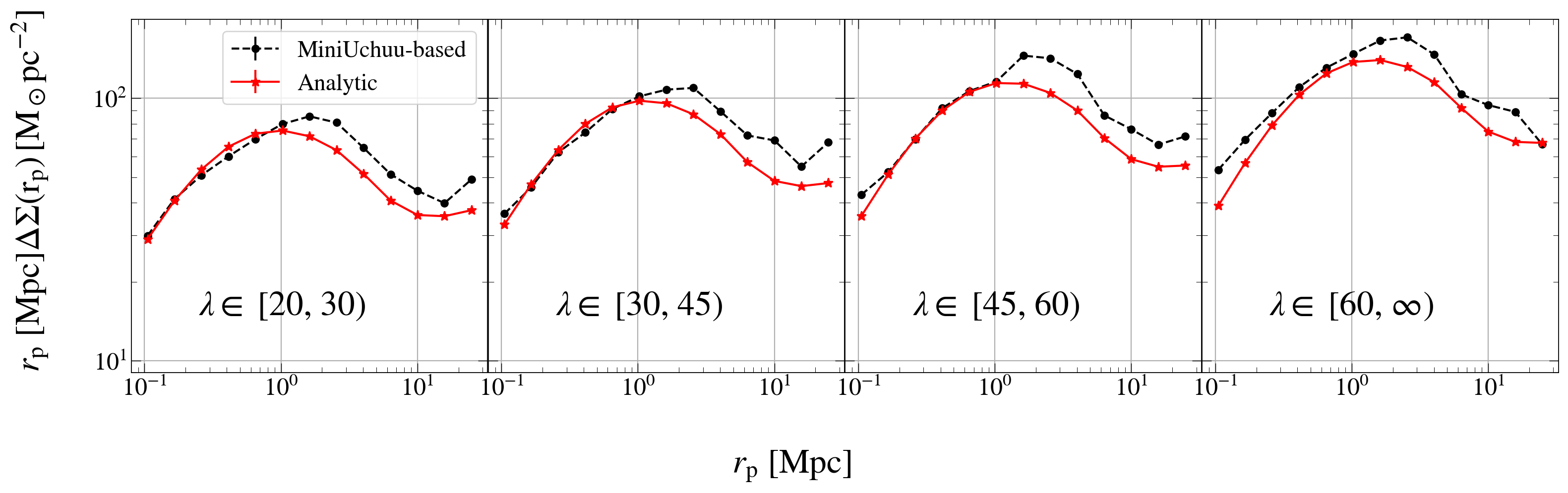}
\caption{Stacked lensing signal derived from the MiniUchuu-based mock (black) in four richness bins at $z=0.3$, with richness modeled by counts-in-cylinders ($\pm 30~\hiMpc$ distance uncertainties).  The red curves show the analytic predictions based on the mean mass and concentration of halos in each bin.  The mean lensing signals of richness-selected clusters are biased high, and the bias is higher on large scales than on small scales. 
} 
\label{fig:cylinder_richness_lensing}
\end{figure*}

\section{Simulations for optical cluster lensing}\label{sec:sims}

We describe the two complementary cluster simulations used in this work: MiniUchuu-based (\ref{sec:MiniUchuu}) and Cardinal (\ref{sec:Cardinal}) mock catalogs.  The former allows us to test a specific projection model, while the latter lets us study realistic red-sequence cluster finding.

Both simulations are built on $N$-body simulations.  The former simulates galaxies using an HOD model, while the latter simulates galaxies using the ADDGALS algorithm \cite{Wechsler2022}.  The former simulates cluster finding using a moderate distance uncertainty of 30~$\hiMpc$, while the latter runs the full \redmapper cluster finder \cite{Rykoff2014, Rykoff2016}, which uses matched filters on galaxy colors, luminosities, and positions to find clusters.   

\subsection{MiniUchuu-based mock catalog}\label{sec:MiniUchuu}

MiniUchuu is an $N$-body simulation from the publicly available Uchuu suite \cite{Ishiyama2021}\footnote{\href{https://www.skiesanduniverses.org/Simulations/Uchuu/}{https://www.skiesanduniverses.org/Simulations/Uchuu/}}, generated with the {\sc 2lptic} initial condition code \cite{Crocce2006} and the {\sc greem} $N$-body code \cite{Ishiyama2009}.  It has $2560^3$ particles within a box size of 400 $\hiMpc$, a dark matter mass resolution of $3.27\times10^8~\hiMsun$, and a force resolution of $4.27~\hikpc$.  The dark matter halos are identified using the Rockstar halo finder \cite{Behroozi2013}.  We focus on the $z=0.3$ snapshot and use the halo mass definition $M_{\rm 200m}$, a spherical overdensity 200 times the mean matter density of the Universe. We only use isolated halos and exclude subhalos.  MiniUchuu is based on a flat $\Lambda$CDM cosmology from {\em Planck} \cite{PlanckCollaboration2020}: 
$\Omega_{\rm m} = 0.3089$, 
$\Omega_{\rm b} = 0.0486$, 
$\sigma_8 = 0.8159$, 
$n_{\rm s} = 0.9667$, and
$h = 0.6774$.

To build the galaxy catalog, we use a simple HOD model \cite{Zheng05} to assign galaxies to halos in MiniUchuu. For halos with mass above $10^{12}\hiMsun$, we assign a central galaxy to the halo center.  For the number of satellite galaxies, we draw an $\Nsat$ value based on a Poisson distribution with mean 
\beq
\langle \Nsat | M \rangle  = 
\left( \frac{M - M_0}{M_1} \right)^{\alpha}~.
\eeq
We use the parameter values from \cite{Sunayama2020}, which are designed to reproduce the DES cluster abundance:
$M_0 = 10^{11.7}~\hiMsun$,  
$M_1 = 10^{12.9}~\hiMsun$, 
and $\alpha=1$. 
The satellite positions are drawn from a spherically symmetric NFW profile \cite{Navarro1997}, with the concentration--mass relation from \cite{Bhattacharya13}. 
We have verified that assigning satellite positions to dark matter particles gives nearly identical results.

The next step is to build a mock cluster catalog from these galaxies.  We assume clusters are centered on dark matter halos with mass above $10^{12.5}~\hiMsun$.  To simulate the observed cluster richness, we use counts-in-cylinders to model the distance uncertainties \cite{Angulo2012, Busch2017, Sunayama2020}; that is, $\lambda$ is the total number of galaxies within a cylinder along the line of sight.  We use a cylinder length of $\pm 30~\hiMpc$, motivated by \cite{Wu2022}. 
This length is likely an underestimate, as recent studies have indicated larger projection depths \cite{Costanzi2019, Myles2021, Lee2025}.  Therefore, our mass bias could be underestimated.
To take into account that richer clusters tend to have larger radii, we determine cluster richness and radius iteratively by solving
\beq
R_\lambda =  \left(\frac{\lambda}{100}\right)^{0.2}
~ {h^{-1}}{\rm Mpc} ~,
\eeq
an optimal relation found in \cite{Rykoff12}. 
If a galaxy is simultaneously contained in the cylinders of multiple halos, we only assign it to the most massive one to avoid double-counting galaxies.

We put our clusters in four richness bins: 
[20, 30), [30, 45), [45, 60), and [60, $\infty$).
For each richness bin, we calculate the cross-correlation between clusters and dark matter particles to derive the lensing signal $\Delta\Sigma$, using the {\tt corrfunc} software \cite{Corrfunc} with a projection depth $\pm~100~\hiMpc$.  This depth is larger than the $\pm~30~\hiMpc$ used for cluster finding.  Increasing this depth further has a negligible effect on the lensing signal.  Figure~\ref{fig:cylinder_richness_lensing} shows the lensing signal derived from MiniUchuu.   In each panel, the black curve shows the MiniUchuu lensing signal in a given richness bin, and the red curve shows the analytical prediction based on the mean mass and concentration of the halos in that bin.  The MiniUchuu results are consistent with the analytic expectation below $\lesssim 1$ Mpc but are higher at larger scales. In Sec.~\ref{sec:mass_bias}, we will show how this biased lensing signal leads to a biased mass calibration.

\subsection{Cardinal}\label{sec:Cardinal}

\begin{figure*}
\centering
\includegraphics[width=2\columnwidth]{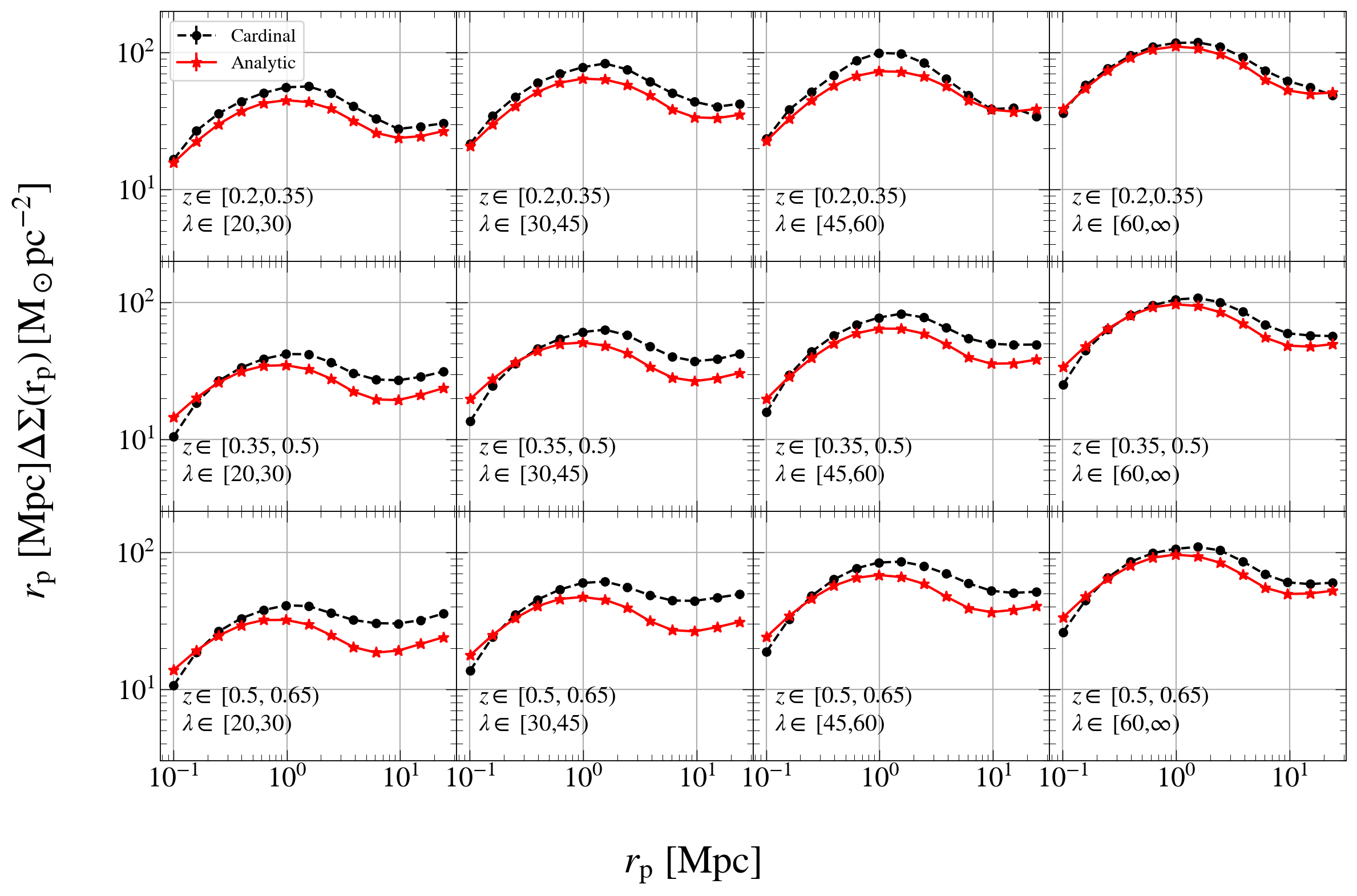}
\caption{Analogous to Fig.~\ref{fig:cylinder_richness_lensing} but for Cardinal, based on the ADDGALS galaxy model and the \redmapper cluster finder. The lensing bias exists on most scales and is stronger than that shown in Fig.~\ref{fig:cylinder_richness_lensing}.
}
\label{fig:cardinal_lensing}
\end{figure*}

Cardinal \cite{To2024} is a suite of lightcone mock catalogs designed to reproduce observed cluster richness and clustering, making it uniquely suitable for our study.  We use the catalog designed for the Dark Energy Survey (DES) with an area of $\approx$5000 deg$^2$. We use the redshift range $0.2 < z < 0.65$, which is constructed from two $N$-body simulations:
The $z<0.315$ lightcone has a mass resolution $3.3\times10^{10}~\hiMsun$ and a force resolution $20~\hikpc$; the $z>0.315$ lightcone has a mass resolution $1.6\times10^{11}~\hiMsun$ and a force resolution $35~\hikpc$. 
We use the virial mass definition based on the $\Delta_{\rm vir}(z)$ from \cite{BryanNorman98} and focus on halos with mass above $10^{13}~\hiMsun$. 
Cardinal is based on a flat $\Lambda$CDM cosmology: 
$\Omega_{\rm m} = 0.286$, 
$\Omega_{\rm b} = 0.047$, 
$\sigma_8 = 0.82$, 
$n_{\rm s} = 0.96$, and
$h = 0.7$.

The galaxy positions, magnitudes, and colors are modeled with the ADDGALS algorithm \cite{Wechsler2022}.  This algorithm calibrates the luminosity--density relation and luminosity--halo mass relation using a high-resolution mock catalog generated with subhalo abundance matching. These relations are then used to populate galaxies in large-volume low-resolution simulations.  Each galaxy is then assigned colors based on its distance to massive halos.  Cardinal \cite{To2024} additionally accounts for tidal stripping in the subhalo abundance matching process, which improves the modeling of galaxies in clusters; as a result, Cardinal can reproduce the observed richness and cluster--galaxy cross-correlation functions.

Because Cardinal includes realistic galaxy colors, the \redmapper cluster finder can be directly applied to it.  The \redmapper algorithm uses the red sequence (the tight color--magnitude relation for galaxies in clusters) to identify clusters and determine their photometric redshifts.  It iteratively calibrates the red-sequence model (the color--magnitude relation as a function of redshift) and identifies clusters \cite{Rykoff2014, Rykoff2016}.
We use the \redmapper catalog with clusters centered on dark matter halos to avoid misidentified cluster centers.  

We put our clusters in 3 redshift bins: 
[0.2,  0.35),
[0.35, 0.5), and
[0.5,  0.65), and 4 richness bins: [20, 30), [30, 45), [45, 60), and [60, $\infty$). For clusters in each bin, we calculate the average of the $\Delta\Sigma$ profiles to simulate the stacked lensing signal. Similar to MiniUchuu, we use the dark matter particles from Cardinal's $N$-body simulations, again using a projection depth of $\pm~100~\hiMpc$.

Figure~\ref{fig:cardinal_lensing} shows the stacked lensing profiles derived from the Cardinal \redmapper catalog.  Similar to the MiniUchuu calculation, we add an analytical model prediction based on the mean mass and concentration in each bin. Compared with MiniUchuu, the Cardinal cluster lensing signal shows a stronger bias on most scales.

\section{Mass calibration in the presence of projection effects}\label{sec:mass_bias}

In this section, we present our mock mass calibration analyses. 
For a redshift and richness bin $k$, we use the following likelihood function to perform the fitting:
\begin{align}
    \ln \mathcal{L}(\Delta\Sigma_k | M, c) \propto -\frac{1}{2} \mathbf{d}_k^{\rm T} \mathcal{C}^{-1} \mathbf{d}_k,
\end{align}
where $\mathbf{d}_k = (\Delta\Sigma_\mathrm{data} - \Delta\Sigma_\mathrm{model})_k$. We use Eq.~(\ref{eq:DeltaSigma}) for $\Delta\Sigma_\mathrm{model}$, and the simulated lensing for $\Delta\Sigma_\mathrm{data}$.  For the covariance matrix $\mathcal{C}$, we use the semi-analytic covariance matrix of DES Year 1 cluster analyses presented in \cite{McClintock2019}, which corresponds to a survey area of 1,437 deg$^2$ and a source density of 6.28 galaxies per arcmin$^{2}$. Using this covariance matrix allows us to assess the mass bias present in the DES cluster mass calibration \cite{McClintock2019}. The statistical uncertainties in the lensing signal will be reduced with larger lens and source samples.

Following the DES analyses \cite{Varga2019, McClintock2019}, for both MiniUchuu and Cardinal, we measure cluster lensing using 15 logarithmically spaced bins between 0.03 and 30 Mpc. For mass fitting, we use the 11 bins between 0.2 and 30 Mpc. To investigate the impact of projection effects on large-scale and small-scale analyses, we split our lensing data vector into small scales, [0.2, 3) Mpc, and large scales, [3, 30) Mpc.   

\begin{figure*}
\centering
\includegraphics[width=1\columnwidth]{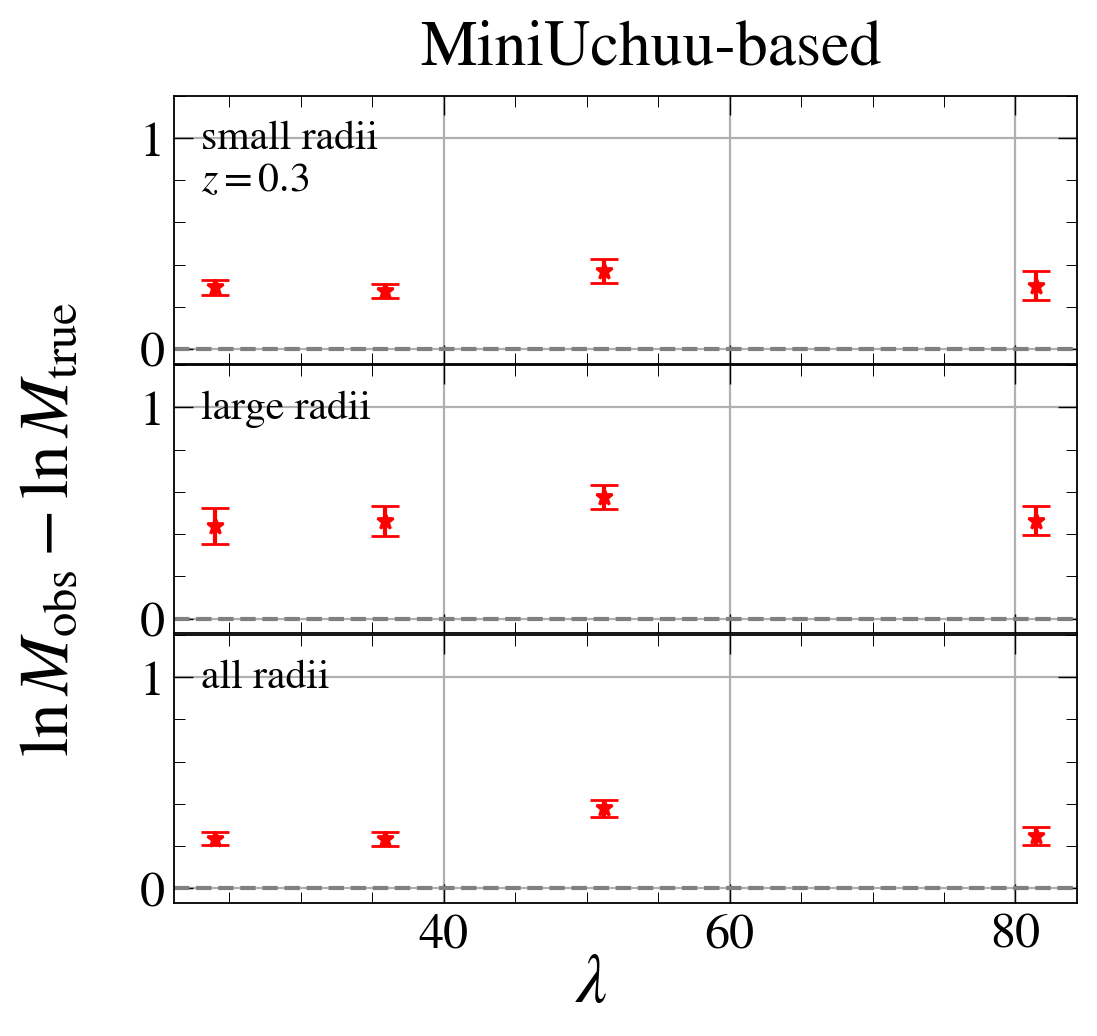}
\includegraphics[width=1\columnwidth]{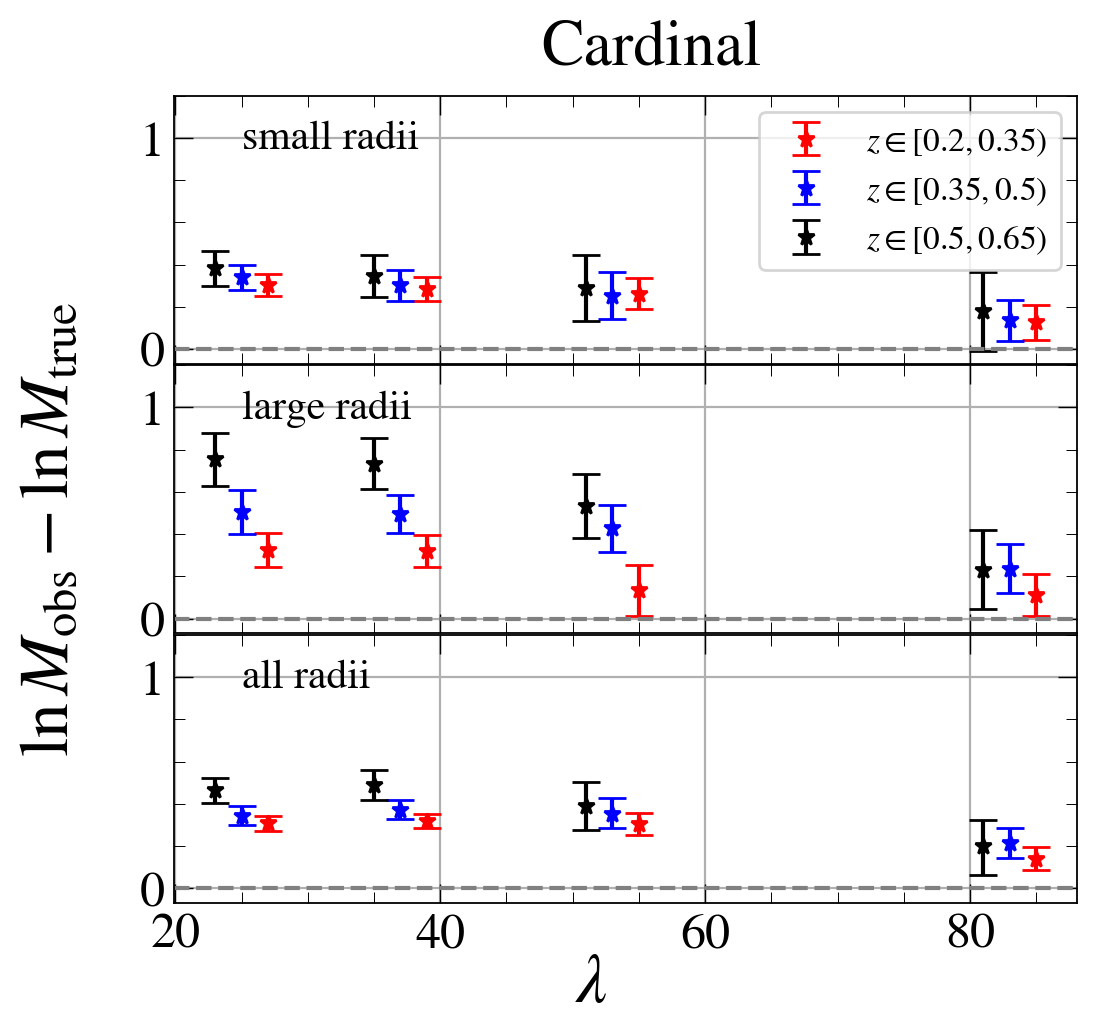}
\caption{Fractional mass bias resulted from projection-induced selection bias, for the MiniUchuu-based mock (left) and Cardinal (right). We fit the analytic model to the mock lensing signals. The points and error bars correspond to the posterior means and 68\% credible intervals.  The top panels use radial range [0.2, 3) Mpc, the middle panels use [3, 30) Mpc, while the bottom panels combine both radial ranges.  Cardinal tends to have larger mass biases than MiniUchuu.  For Cardinal, the bias is higher for high-redshift and low-richness clusters.} 
\label{fig:mass_bias}
\end{figure*}

We use two free parameters, $\log_{10} M$ and $c$, in our fitting, and we assume flat priors, $\log_{10} M\in(12, 16)$ and $c\in(2, 10)$.  We use the \textit{emcee} software \cite{Foreman-Mackey2013} for the Markov chain Monte Carlo analyses, using 32 walkers for 10,000 iterations and discarding the first 1,000 iterations as burn-in.  Figure~\ref{fig:mass_bias} shows the difference in the posterior mean mass $M_{\rm obs}$ and the true mean mass $M_{\rm true}$.  The error bars show the 68\% credible interval.  We show the difference in the natural logarithm, which can be interpreted as a fractional difference. 

The left-hand panel of Fig.~\ref{fig:mass_bias} shows the MiniUchuu results.  For small scales, [0.2, 3) Mpc, the mass is biased high by $\approx25\%$.  For large scales, [3, 30) Mpc, the bias is $\approx50\%$.  Since the small-scale lensing dominates the statistical power, combining all scales gives a bias of $\approx25\%$, similar to small scales. For MiniUchuu, the mass bias is nearly independent of richness. The lack of richness dependence in lensing bias is also seen in \cite{Sunayama2020} using an HOD-based mock catalog (see their Fig.~8).  This appears contradictory to Refs.~\cite{Myles2021, Lee2025, Myles25projection}, which find that the fraction of projected galaxies ($f_{\rm proj}$) is higher for low-richness clusters.  However, a higher $f_{\rm proj}$ does not necessarily lead to a higher lensing bias, because the lensing bias depends on the coherent shift in richness and lensing at a given halo mass.

The right-hand panel of Fig.~\ref{fig:mass_bias} shows the Cardinal results, which have a larger bias than MiniUchuu and show a richness trend.  For small scales, the bias is $20-40\%$; for large scales, the bias is $20-80\%$.  Combining all scales gives a bias of $20-50\%$, similar to the small-scale analysis.  The richness and redshift trend is the strongest for the large-scale analysis, with an 80\% bias at $z\geq0.5$ and $\lambda\approx20$. At high redshift, the selection bias is stronger due to the larger photometric redshift uncertainties.

MiniUchuu and Cardinal differ in their cosmological parameters, but they have similar $\sigma_8$ values (0.8159 vs.~0.82).  Therefore, the difference in mass bias is not due to different matter clustering but rather different galaxy--halo connection and cluster finding. We note that both mock catalogs have limitations. The HOD model used for MiniUchuu could be too restrictive and lacks environmental dependence.  The Cardinal mock is based on a relatively low-resolution $N$-body simulation, and the detailed red galaxy distribution depends on the local density estimator and a tidal stripping model.

\section{Summary and discussion}\label{sec:summary}

A cluster sample selected by optical richness tends to have a mean lensing signal higher than expected from its mean mass, and this bias is caused by the coherent boost of richness and lensing due to projection effects.  To assess the impact of this selection bias on cluster mass calibration, we have performed mock mass calibrations using two simulations with different cluster finders: The MiniUchuu-based mock simulates a 30 $\hiMpc$ distance uncertainty using counts-in-cylinders, and the Cardinal mock uses the \redmapper cluster finder.  
We have found a $25\%$ bias in the former and $20-50\%$ in the latter, and that large-scale-focused mass calibration ($\rp>3$ Mpc) has a larger bias compared with the small-scale-focused mass calibration.  The Cardinal results show that the bias is larger for high-redshift and low-richness clusters.

Our findings confirm that the projection-induced selection bias is the most detrimental of the known systematic biases impacting cluster mass calibration.  
Other known systematics include the impact of baryons on the mass distribution \cite{Schaller15, Henson17, Lee18, Cromer21, GiriSchneider21, Grandis21b, Sunayama23,  To24baryon, Dalal25}, the offset between the inferred cluster centers and the true halo centers (miscentering) \cite{Johnston07, Kohlinger15, Zhang2019, Sommer22}, and the diluted lensing signal due to cluster members (the so-called boost factor) \cite{Kohlinger15, Varga2019}.  Ref.~\cite{Bocquet24a} has assessed the impact of these systematics on the weak lensing mass bias for $\sim1000$ clusters selected in millimeter-wave by the South Pole Telescope, which is unaffected by the optical selection bias discussed in this paper.  Their weak lensing mass bias is dominated by baryonic effects (for $z \lesssim 0.4$) and the photometric redshift calibration of source galaxies (for $z \gtrsim 0.4$); see their Table IV and Fig.~10.  These systematics lead to a mass bias around $5-10\%$; their highest bias is 16\%, which occurs at $z \approx 1$.   These systematics also impact optically selected clusters, but they are subdominant compared with the optical selection bias discussed in this paper.

How do we constrain the projection-induced selection bias using observational data?  Spectroscopic redshifts of member galaxies can help us separate projected vs.~physically-associated cluster members \cite{Myles2021, Myles25projection}.  Cross-correlating clusters with spectroscopic galaxies can help us calibrate the redshift distribution of cluster members \cite{Sunayama23, Yang25}. Ref.~\cite{Costanzi2019} also shows how to use photometric galaxies to calibrate projection effects. Combining cluster lensing and cluster--galaxy cross correlations can help us self-calibrate the selection bias, which boosts all cluster-related 2-point statistics \cite{To21a, To2021, Sunayama23, Zeng2023}.  Multiwavelength analyses comparing cluster lensing signals of optically selected clusters and gas-selected clusters (observed in mm-wave or x-ray) can help us calibrate the lensing bias \cite{Zhou24}. In sum, we advocate a multiprobe, multiwavelength approach for cluster cosmology. Future modeling strategies should focus on self-consistently modeling the impact of projection on richness, lensing, and gas observables \cite{Angulo2012, Shirasaki16, Ragagnin25}.

How do we treat the selection bias in optical cluster cosmology analyses?  Recent studies have modeled the selection bias as functions of radius, richness, and redshift \cite{To2021, Sunayama23, DESY3CL, To25}.
Another promising route is to use $N$-body simulations to forward-model richness and lensing, an approach that automatically takes into account the selection bias \cite{Salcedo2024, Lee2025, Cao26unify}. Since the mass bias depends on the galaxy model, the cluster finding, and the radial range used, we will need a flexible model, which can be achieved with simulation-based forward modeling.  We anticipate that the simulation-based forward modeling has the potential to help us realize the constraining power of optical clusters.

\begin{acknowledgments}

This work has been supported by DOE Grants DE-SC0021916 and DE-SC0010129, NASA Grant 15-WFIRST15-0008, and NSF Grant AST-2509910.  AT~is supported by the Spanish National Research Council (CSIC) through grant No.~20225AT025. We thank Boise State University's Research Computing Department for providing high-performance computing support for the Borah computer cluster (DOI: 10.18122/oit/3/boisestate).

\end{acknowledgments}


\bibliographystyle{apsrev4-2-author-truncate}
\bibliography{references, master_refs2}

@ARTICLE{Cao26unify,
       author = {{Cao}, Shulei and {Wu}, Hao-Yi and {Salcedo}, Andr{\'e}s N. and {Weinberg}, David H. and {Schaller}, Matthieu and {Schaye}, Joop},
        title = "{Unifying cluster and galaxy cosmology analyses using the galaxy-halo connection}",
      journal = {arXiv e-prints},
         year = 2026,
        month = jan,
          eid = {arXiv:2601.16353},
        pages = {arXiv:2601.16353},
          doi = {10.48550/arXiv.2601.16353},
archivePrefix = {arXiv},
       eprint = {2601.16353},
 primaryClass = {astro-ph.CO},
       adsurl = {https://ui.adsabs.harvard.edu/abs/2026arXiv260116353C}
}

@ARTICLE{Yang25,
       author = {{Yang}, Lei and {Wu}, Hao-Yi and others},
        title = "{}",
      journal = {arXiv e-prints},
         year = 2025,
        month = oct,
          eid = {in prep},
        pages = {arXiv:},
          doi = {},
archivePrefix = {arXiv},
       eprint = {},
 primaryClass = {astro-ph.CO},
       adsurl = {}
}

@ARTICLE{Dalal25,
       author = {{Dalal}, Nihar and {To}, Chun-Hao and {Hirata}, Chris and {Hyeon-Shin}, Tae and {Hilton}, Matt and {Pandey}, Shivam and {Bond}, J. Richard},
        title = "{Deciphering Baryonic Feedback from ACT tSZ Galaxy Clusters}",
      journal = {arXiv e-prints},
         year = 2025,
        month = jul,
          eid = {arXiv:2507.04476},
        pages = {arXiv:2507.04476},
          doi = {10.48550/arXiv.2507.04476},
archivePrefix = {arXiv},
       eprint = {2507.04476},
 primaryClass = {astro-ph.CO},
       adsurl = {https://ui.adsabs.harvard.edu/abs/2025arXiv250704476D}
}

@ARTICLE{To24baryon,
       author = {{To}, Chun-Hao and {Pandey}, Shivam and {Krause}, Elisabeth and {Dalal}, Nihar and {Anbajagane}, Dhayaa and {Weinberg}, David H.},
        title = "{Deciphering baryonic feedback with galaxy clusters}",
      journal = {\jcap},
         year = 2024,
        month = jul,
       volume = {2024},
       number = {7},
          eid = {037},
        pages = {037},
          doi = {10.1088/1475-7516/2024/07/037},
archivePrefix = {arXiv},
       eprint = {2402.00110},
 primaryClass = {astro-ph.CO},
       adsurl = {https://ui.adsabs.harvard.edu/abs/2024JCAP...07..037T}
}

@ARTICLE{Pratt19,
       author = {{Pratt}, G.~W. and {Arnaud}, M. and {Biviano}, A. and {Eckert}, D. and {Ettori}, S. and {Nagai}, D. and {Okabe}, N. and {Reiprich}, T.~H.},
        title = "{The Galaxy Cluster Mass Scale and Its Impact on Cosmological Constraints from the Cluster Population}",
      journal = {\ssr},
         year = 2019,
        month = feb,
       volume = {215},
       number = {2},
          eid = {25},
        pages = {25},
          doi = {10.1007/s11214-019-0591-0},
archivePrefix = {arXiv},
       eprint = {1902.10837},
 primaryClass = {astro-ph.CO},
       adsurl = {https://ui.adsabs.harvard.edu/abs/2019SSRv..215...25P}
}

@ARTICLE{Takahashi12halofit,
       author = {{Takahashi}, Ryuichi and {Sato}, Masanori and {Nishimichi}, Takahiro and {Taruya}, Atsushi and {Oguri}, Masamune},
        title = "{Revising the Halofit Model for the Nonlinear Matter Power Spectrum}",
      journal = {\apj},
         year = 2012,
        month = dec,
       volume = {761},
       number = {2},
          eid = {152},
        pages = {152},
          doi = {10.1088/0004-637X/761/2/152},
archivePrefix = {arXiv},
       eprint = {1208.2701},
 primaryClass = {astro-ph.CO},
       adsurl = {https://ui.adsabs.harvard.edu/abs/2012ApJ...761..152T}
}

@ARTICLE{Ragagnin25,
       author = {{Euclid Collaboration} and {Ragagnin}, A. and {Saro}, A. and {Andreon}, S. and {Biviano}, A. and {Dolag}, K. and {Ettori}, S. and {Giocoli}, C. and {Le Brun}, A.~M.~C. and {Mamon}, G.~A. and {Maughan}, B.~J. and {Meneghetti}, M. and {Moscardini}, L. and {Pacaud}, F. and {Pratt}, G.~W. and {Sereno}, M. and {Borgani}, S. and {Calura}, F. and {Castignani}, G. and {De Petris}, M. and {Eckert}, D. and {Lesci}, G.~F. and {Macias-Perez}, J. and {Maturi}, M. and others},
        title = "{Euclid preparation: LXVI. Impact of line-of-sight projections on the covariance between galaxy cluster multi-wavelength observable properties: insights from hydrodynamic simulations}",
      journal = {\aap},
         year = 2025,
        month = mar,
       volume = {695},
          eid = {A282},
        pages = {A282},
          doi = {10.1051/0004-6361/202451347},
archivePrefix = {arXiv},
       eprint = {2412.00191},
 primaryClass = {astro-ph.CO},
       adsurl = {https://ui.adsabs.harvard.edu/abs/2025A&A...695A.282E}
}

@ARTICLE{Sommer22,
       author = {{Sommer}, Martin W. and {Schrabback}, Tim and {Applegate}, Douglas E. and {Hilbert}, Stefan and {Ansarinejad}, Behzad and {Floyd}, Benjamin and {Grandis}, Sebastian},
        title = "{Weak lensing mass modeling bias and the impact of miscentring}",
      journal = {\mnras},
         year = 2022,
        month = jan,
       volume = {509},
       number = {1},
        pages = {1127-1146},
          doi = {10.1093/mnras/stab3052},
archivePrefix = {arXiv},
       eprint = {2105.08027},
 primaryClass = {astro-ph.CO},
       adsurl = {https://ui.adsabs.harvard.edu/abs/2022MNRAS.509.1127S}
}

@ARTICLE{Zhou24,
       author = {{Zhou}, Conghao and {Wu}, Hao-Yi and {Salcedo}, Andr{\'e}s N. and {Grandis}, Sebastian and {Jeltema}, Tesla and {Leauthaud}, Alexie and {Costanzi}, Matteo and {Sunayama}, Tomomi and {Weinberg}, David H. and {Zhang}, Tianyu and {Rozo}, Eduardo and {To}, Chun-Hao and {Bocquet}, Sebastian and {Varga}, Tamas and {Kwiecien}, Matthew},
        title = "{Forecasting the constraints on optical selection bias and projection effects of galaxy cluster lensing with multiwavelength data}",
      journal = {\prd},
         year = 2024,
        month = nov,
       volume = {110},
       number = {10},
          eid = {103508},
        pages = {103508},
          doi = {10.1103/PhysRevD.110.103508},
archivePrefix = {arXiv},
       eprint = {2312.11789},
 primaryClass = {astro-ph.CO},
       adsurl = {https://ui.adsabs.harvard.edu/abs/2024PhRvD.110j3508Z}
}

@ARTICLE{DESY3CL,
       author = {{DES Collaboration} and {Abbott}, T.~M.~C. and {Aguena}, M. and {Alarcon}, A. and {Anbajagane}, D. and {Andrade-Oliveira}, F. and {Avila}, S. and {Bacon}, D. and {Becker}, M.~R. and {Bhargava}, S. and {Blazek}, J. and {Bocquet}, S. and {Brooks}, D. and {Carnero Rosell}, A. and {Carretero}, J. and {Castander}, F.~J. and {Chang}, C. and {Choi}, A. and {Conselice}, C. and {Costanzi}, M. and {Crocce}, M. and others},
        title = "{Dark Energy Survey Year 3 Results: Cosmological Constraints from Cluster Abundances, Weak Lensing, and Galaxy Clustering}",
      journal = {arXiv e-prints},
         year = 2025,
        month = mar,
          eid = {arXiv:2503.13632},
        pages = {arXiv:2503.13632},
          doi = {10.48550/arXiv.2503.13632},
archivePrefix = {arXiv},
       eprint = {2503.13632},
 primaryClass = {astro-ph.CO},
       adsurl = {https://ui.adsabs.harvard.edu/abs/2025arXiv250313632D}
}

@ARTICLE{To25,
       author = {{To}, Chun-Hao and {Krause}, Elisabeth and {Chang}, Chihway and {Wu}, Hao-Yi and {Wechsler}, Risa H. and {Rozo}, Eduardo and {Weinberg}, David H. and {Anbajagane}, D. and {Avila}, S. and {Blazek}, J. and {Bocquet}, S. and {Costanzi}, M. and {De Vicente}, J. and {Elvin-Poole}, J. and {Fert{\'e}}, A. and {Grandis}, S. and {Muir}, J. and {Porredon}, A. and {Samuroff}, S. and {Sanchez}, E. and {Cid}, D. Sanchez and {Sevilla-Noarbe}, I. and {Weaverdyck}, N. and {Abbott}, T.~M.~C. and {Aguena}, M. and {Andrade-Oliveira}, F. and {Bacon}, D. and {Becker}, M.~R. and {Brooks}, D. and {Rosell}, A. Carnero and {Carretero}, J. and {Choi}, A. and {da Costa}, L.~N. and {Pereira}, M.~E.~S. and {Davis}, T.~M. and {Desai}, S. and {Doel}, P. and {Doux}, C. and {Everett}, S. and {Frieman}, J. and {Garc{\'\i}a-Bellido}, J. and {Gatti}, M. and {Gaztanaga}, E. and {Giannini}, G. and {Gruen}, D. and {Gutierrez}, G. and {Hinton}, S.~R. and {Hollowood}, D.~L. and {Honscheid}, K. and {Jeltema}, T. and {Kuehn}, K. and {Lee}, S. and {Marshall}, J.~L. and {Mena-Fern{\'a}ndez}, J. and {Miquel}, R. and {Mohr}, J.~J. and {Myles}, J. and {Palmese}, A. and {Malag{\'o}n}, A.~A. Plazas and {Romer}, A.~K. and {Shin}, T. and {Smith}, M. and {Suchyta}, E. and {Tarle}, G. and {Vikram}, V. and {Walker}, A.~R. and {Weller}, J. and {DES Collaboration}},
        title = "{Dark energy survey: Modeling strategy for multiprobe cluster cosmology and validation for the full six-year dataset}",
      journal = {\prd},
         year = 2025,
        month = sep,
       volume = {112},
       number = {6},
          eid = {063537},
        pages = {063537},
          doi = {10.1103/ynqj-6hsb},
archivePrefix = {arXiv},
       eprint = {2503.13631},
 primaryClass = {astro-ph.CO},
       adsurl = {https://ui.adsabs.harvard.edu/abs/2025PhRvD.112f3537T}
}

@ARTICLE{Bocquet24a,
       author = {{Bocquet}, S. and {Grandis}, S. and {Bleem}, L.~E. and {Klein}, M. and {Mohr}, J.~J. and {Aguena}, M. and {Alarcon}, A. and {Allam}, S. and {Allen}, S.~W. and {Alves}, O. and {Amon}, A. and {Ansarinejad}, B. and {Bacon}, D. and {Bayliss}, M. and {Bechtol}, K. and {Becker}, M.~R. and {Benson}, B.~A. and {Bernstein}, G.~M. and {Brodwin}, M. and {Brooks}, D. and {Campos}, A. and {Canning}, R.~E.~A. and {Carlstrom}, J.~E. and {Carnero Rosell}, A. and {Carrasco Kind}, M. and {Carretero}, J. and {Cawthon}, R. and {Chang}, C. and {Chen}, R. and {Choi}, A. and {Cordero}, J. and {Costanzi}, M. and {da Costa}, L.~N. and {Pereira}, M.~E.~S. and others},
        title = "{SPT clusters with DES and HST weak lensing. I. Cluster lensing and Bayesian population modeling of multiwavelength cluster datasets}",
      journal = {\prd},
         year = 2024,
        month = oct,
       volume = {110},
       number = {8},
          eid = {083509},
        pages = {083509},
          doi = {10.1103/PhysRevD.110.083509},
archivePrefix = {arXiv},
       eprint = {2310.12213},
 primaryClass = {astro-ph.CO},
       adsurl = {https://ui.adsabs.harvard.edu/abs/2024PhRvD.110h3509B}
}

@ARTICLE{Sunayama23,
       author = {{Sunayama}, Tomomi},
        title = "{Observational constraints of an anisotropic boost due to the projection effects using redMaPPer clusters}",
      journal = {\mnras},
         year = 2023,
        month = jun,
       volume = {521},
       number = {4},
        pages = {5064-5076},
          doi = {10.1093/mnras/stad786},
       adsurl = {https://ui.adsabs.harvard.edu/abs/2023MNRAS.521.5064S}
}

@ARTICLE{YooSeljak12,
       author = {{Yoo}, Jaiyul and {Seljak}, Uro{\v{s}}},
        title = "{Joint analysis of gravitational lensing, clustering, and abundance: Toward the unification of large-scale structure analysis}",
      journal = {\prd},
         year = 2012,
        month = oct,
       volume = {86},
       number = {8},
          eid = {083504},
        pages = {083504},
          doi = {10.1103/PhysRevD.86.083504},
archivePrefix = {arXiv},
       eprint = {1207.2471},
 primaryClass = {astro-ph.CO},
       adsurl = {https://ui.adsabs.harvard.edu/abs/2012PhRvD..86h3504Y}
}

@ARTICLE{Grandis21b,
       author = {{Grandis}, Sebastian and {Bocquet}, Sebastian and {Mohr}, Joseph J. and {Klein}, Matthias and {Dolag}, Klaus},
        title = "{Calibration of bias and scatter involved in cluster mass measurements using optical weak gravitational lensing}",
      journal = {\mnras},
         year = 2021,
        month = nov,
       volume = {507},
       number = {4},
        pages = {5671-5689},
          doi = {10.1093/mnras/stab2414},
archivePrefix = {arXiv},
       eprint = {2103.16212},
 primaryClass = {astro-ph.CO},
       adsurl = {https://ui.adsabs.harvard.edu/abs/2021MNRAS.507.5671G}
}

@ARTICLE{Lee18,
       author = {{Lee}, B.~E. and {Le Brun}, A.~M.~C. and {Haq}, M.~E. and {Deering}, N.~J. and {King}, L.~J. and {Applegate}, D. and {McCarthy}, I.~G.},
        title = "{The relative impact of baryons and cluster shape on weak lensing mass estimates of galaxy clusters}",
      journal = {\mnras},
         year = 2018,
        month = sep,
       volume = {479},
       number = {1},
        pages = {890-899},
          doi = {10.1093/mnras/sty1377},
archivePrefix = {arXiv},
       eprint = {1805.12186},
 primaryClass = {astro-ph.CO},
       adsurl = {https://ui.adsabs.harvard.edu/abs/2018MNRAS.479..890L}
}

@ARTICLE{Myles25projection,
       author = {{Myles}, J. and {Gruen}, D. and {Jeltema}, T. and {Fu}, S. and {Kremin}, A. and {Aguilar}, J. and {Ahlen}, S. and {Bianchi}, D. and {Brooks}, D. and {Castander}, F.~J. and {Claybaugh}, T. and {de la Macorra}, A. and {Dey}, A. and {Doel}, P. and {Ferraro}, S. and {Forero-Romero}, J.~E. and {Gazta{\~n}aga}, E. and {Gontcho}, S. Gontcho A and {Gutierrez}, G. and {Honscheid}, K. and {Ishak}, M. and {Kehoe}, R. and {Kirkby}, D. and {Kisner}, T. and {Lahav}, O. and {Landriau}, M. and {LeGuillou}, L. and {Manera}, M. and {Meisner}, A. and {Miquel}, R. and {Moustakas}, J. and {Nadathur}, S. and {Newman}, J.~A. and {Palanque-Delabrouille}, N. and {Prada}, F. and {P{\'e}rez-R{\`a}fols}, I. and {Rossi}, G. and {Sanchez}, E. and {Schlegel}, D. and {Schubnell}, M. and {Silber}, J. and {Sprayberry}, D. and {Tarl{\'e}}, G. and {Weaver}, B.~A. and {Zhou}, R.},
        title = "{Spectroscopic Characterization of redMaPPer Galaxy Clusters with DESI}",
      journal = {arXiv e-prints},
         year = 2025,
        month = jun,
          eid = {arXiv:2506.06249},
        pages = {arXiv:2506.06249},
          doi = {10.48550/arXiv.2506.06249},
archivePrefix = {arXiv},
       eprint = {2506.06249},
 primaryClass = {astro-ph.CO},
       adsurl = {https://ui.adsabs.harvard.edu/abs/2025arXiv250606249M}
}

@ARTICLE{Zhang23triaxiality,
       author = {{Zhang}, Zhuowen and {Wu}, Hao-Yi and {Zhang}, Yuanyuan and {Frieman}, Joshua and {To}, Chun-Hao and {DeRose}, Joseph and {Costanzi}, Matteo and {Wechsler}, Risa H. and {Adhikari}, Susmita and {Rykoff}, Eli and {Jeltema}, Tesla and {Evrard}, August and {Rozo}, Eduardo},
        title = "{Modelling galaxy cluster triaxiality in stacked cluster weak lensing analyses}",
      journal = {\mnras},
         year = 2023,
        month = aug,
       volume = {523},
       number = {2},
        pages = {1994-2013},
          doi = {10.1093/mnras/stad1404},
archivePrefix = {arXiv},
       eprint = {2202.08211},
 primaryClass = {astro-ph.CO},
       adsurl = {https://ui.adsabs.harvard.edu/abs/2023MNRAS.523.1994Z}
}

@ARTICLE{Wu21,
       author = {{Wu}, Hao-Yi and {Weinberg}, David H. and {Salcedo}, Andr{\'e}s N. and {Wibking}, Benjamin D.},
        title = "{Cosmology with Galaxy Cluster Weak Lensing: Statistical Limits and Experimental Design}",
      journal = {\apj},
         year = 2021,
        month = mar,
       volume = {910},
       number = {1},
          eid = {28},
        pages = {28},
          doi = {10.3847/1538-4357/abdc23},
archivePrefix = {arXiv},
       eprint = {2012.01956},
 primaryClass = {astro-ph.CO},
       adsurl = {https://ui.adsabs.harvard.edu/abs/2021ApJ...910...28W}
}

@ARTICLE{Akeson19,
       author = {{Akeson}, Rachel and {Armus}, Lee and {Bachelet}, Etienne and
         {Bailey}, Vanessa and {Bartusek}, Lisa and {Bellini}, Andrea and
         {Benford}, Dominic and {Bennett}, David and {Bhattacharya}, Aparna and
         {Bohlin}, Ralph and {Boyer}, Martha and {Bozza}, Valerio and
         {Bryden}, Geoffrey and {Calchi Novati}, Sebastiano and
         {Carpenter}, Kenneth and {Casertano}, Stefano and {Choi}, Ami and
         {Content}, David and {Dayal}, Pratika and {Dressler}, Alan and
         {Dor{\'e}}, Olivier and {Fall}, S. Michael and {Fan}, Xiaohui and
         {Fang}, Xiao and {Filippenko}, Alexei and {Finkelstein}, Steven and
         {Foley}, Ryan and {Furlanetto}, Steven and {Kalirai}, Jason and
         {Gaudi}, B. Scott and {Gilbert}, Karoline and {Girard}, Julien and
         {Grady}, Kevin and {Greene}, Jenny and {Guhathakurta}, Puragra and
         {Heinrich}, Chen and {Hemmati}, Shoubaneh and {Hendel}, David and
         {Henderson}, Calen and {Henning}, Thomas and {Hirata}, Christopher and
         {Ho}, Shirley and {Huff}, Eric and {Hutter}, Anne and {Jansen}, Rolf and
         {Jha}, Saurabh and {Johnson}, Samson and {Jones}, David and
         {Kasdin}, Jeremy and {Kelly}, Patrick and {Kirshner}, Robert and
         {Koekemoer}, Anton and {Kruk}, Jeffrey and {Lewis}, Nikole and
         {Macintosh}, Bruce and {Madau}, Piero and {Malhotra}, Sangeeta and {Mand
        el}, Kaisey and {Massara}, Elena and {Masters}, Daniel and
         {McEnery}, Julie and {McQuinn}, Kristen and {Melchior}, Peter and
         {Melton}, Mark and {Mennesson}, Bertrand and {Peeples}, Molly and
         {Penny}, Matthew and {Perlmutter}, Saul and {Pisani}, Alice and
         {Plazas}, Andr{\'e}s and {Poleski}, Radek and {Postman}, Marc and
         {Ranc}, Cl{\'e}ment and {Rauscher}, Bernard and {Rest}, Armin and
         {Roberge}, Aki and {Robertson}, Brant and {Rodney}, Steven and
         {Rhoads}, James and {Rhodes}, Jason and {Ryan}, Russell, Jr. and
         {Sahu}, Kailash and {Sand}, David and {Scolnic}, Dan and {Seth}, Anil and
         {Shvartzvald}, Yossi and {Siellez}, Karelle and {Smith}, Arfon and
         {Spergel}, David and {Stassun}, Keivan and {Street}, Rachel and
         {Strolger}, Louis-Gregory and {Szalay}, Alexander and {Trauger}, John and
         {Troxel}, M.~A. and {Turnbull}, Margaret and {van der Marel}, Roeland and
         {von der Linden}, Anja and {Wang}, Yun and {Weinberg}, David and
         {Williams}, Benjamin and {Windhorst}, Rogier and {Wollack}, Edward and
         {Wu}, Hao-Yi and {Yee}, Jennifer and {Zimmerman}, Neil},
        title = "{The Wide Field Infrared Survey Telescope: 100 Hubbles for the 2020s}",
      journal = {arXiv e-prints},
         year = 2019,
        month = feb,
          eid = {arXiv:1902.05569},
        pages = {arXiv:1902.05569},
archivePrefix = {arXiv},
       eprint = {1902.05569},
 primaryClass = {astro-ph.IM},
       adsurl = {https://ui.adsabs.harvard.edu/abs/2019arXiv190205569A}
}

@ARTICLE{To21a,
       author = {{To}, Chun-Hao and {Krause}, Elisabeth and {Rozo}, Eduardo and {Wu}, Hao-Yi and {Gruen}, Daniel and {DeRose}, Joseph and {Rykoff}, Eli and {Wechsler}, Risa H. and {Becker}, Matthew and {Costanzi}, Matteo and {Eifler}, Tim and {da Silva Pereira}, Maria Elidaiana and {Kokron}, Nickolas and {DES Collaboration}},
        title = "{Combination of cluster number counts and two-point correlations: validation on mock Dark Energy Survey}",
      journal = {\mnras},
         year = 2021,
        month = apr,
       volume = {502},
       number = {3},
        pages = {4093-4111},
          doi = {10.1093/mnras/stab239},
archivePrefix = {arXiv},
       eprint = {2008.10757},
 primaryClass = {astro-ph.CO},
       adsurl = {https://ui.adsabs.harvard.edu/abs/2021MNRAS.502.4093T}
}

@ARTICLE{Cromer21,
       author = {{Cromer}, Dylan and {Battaglia}, Nicholas and {Miyatake}, Hironao and {Simet}, Melanie},
        title = "{Towards 1\% accurate galaxy cluster masses: including baryons in weak-lensing mass inference}",
      journal = {\jcap},
         year = 2022,
        month = oct,
       volume = {2022},
       number = {10},
          eid = {034},
        pages = {034},
          doi = {10.1088/1475-7516/2022/10/034},
archivePrefix = {arXiv},
       eprint = {2104.06925},
 primaryClass = {astro-ph.CO},
       adsurl = {https://ui.adsabs.harvard.edu/abs/2022JCAP...10..034C}
}

@ARTICLE{Henson17,
       author = {{Henson}, Monique A. and {Barnes}, David J. and {Kay}, Scott T. and
         {McCarthy}, Ian G. and {Schaye}, Joop},
        title = "{The impact of baryons on massive galaxy clusters: halo structure and cluster mass estimates}",
      journal = {\mnras},
         year = 2017,
        month = mar,
       volume = {465},
       number = {3},
        pages = {3361-3378},
          doi = {10.1093/mnras/stw2899},
archivePrefix = {arXiv},
       eprint = {1607.08550},
 primaryClass = {astro-ph.CO},
       adsurl = {https://ui.adsabs.harvard.edu/abs/2017MNRAS.465.3361H}
}

@ARTICLE{Eifler21HLS,
       author = {{Eifler}, Tim and {Miyatake}, Hironao and {Krause}, Elisabeth and {Heinrich}, Chen and {Miranda}, Vivian and {Hirata}, Christopher and {Xu}, Jiachuan and {Hemmati}, Shoubaneh and {Simet}, Melanie and {Capak}, Peter and {Choi}, Ami and {Dor{\'e}}, Olivier and {Doux}, Cyrille and {Fang}, Xiao and {Hounsell}, Rebekah and {Huff}, Eric and {Huang}, Hung-Jin and {Jarvis}, Mike and {Kruk}, Jeffrey and {Masters}, Dan and {Rozo}, Eduardo and {Scolnic}, Dan and {Spergel}, David N. and {Troxel}, Michael and {von der Linden}, Anja and {Wang}, Yun and {Weinberg}, David H. and {Wenzl}, Lukas and {Wu}, Hao-Yi},
        title = "{Cosmology with the Roman Space Telescope - multiprobe strategies}",
      journal = {\mnras},
         year = 2021,
        month = oct,
       volume = {507},
       number = {2},
        pages = {1746-1761},
          doi = {10.1093/mnras/stab1762},
archivePrefix = {arXiv},
       eprint = {2004.05271},
 primaryClass = {astro-ph.CO},
       adsurl = {https://ui.adsabs.harvard.edu/abs/2021MNRAS.507.1746E}
}

@ARTICLE{DESY1CL,
       author = {{DES Collaboration}},
        title = "{Dark Energy Survey Year 1 Results: Cosmological constraints from cluster abundances and weak lensing}",
      journal = {\prd},
         year = 2020,
        month = jul,
       volume = {102},
       number = {2},
          eid = {023509},
        pages = {023509},
          doi = {10.1103/PhysRevD.102.023509},
archivePrefix = {arXiv},
       eprint = {2002.11124},
 primaryClass = {astro-ph.CO},
       adsurl = {https://ui.adsabs.harvard.edu/abs/2020PhRvD.102b3509A}
}

@ARTICLE{Osato18,
       author = {{Osato}, Ken and {Nishimichi}, Takahiro and {Oguri}, Masamune and
         {Takada}, Masahiro and {Okumura}, Teppei},
        title = "{Strong orientation dependence of surface mass density profiles of dark haloes at large scales}",
      journal = {\mnras},
         year = "2018",
        month = "Jun",
       volume = {477},
       number = {2},
        pages = {2141-2153},
          doi = {10.1093/mnras/sty762},
archivePrefix = {arXiv},
       eprint = {1712.00094},
 primaryClass = {astro-ph.CO},
       adsurl = {https://ui.adsabs.harvard.edu/abs/2018MNRAS.477.2141O}
}

@ARTICLE{Kohlinger15,
       author = {{K{\"o}hlinger}, F. and {Hoekstra}, H. and {Eriksen}, M.},
        title = "{Statistical uncertainties and systematic errors in weak lensing mass estimates of galaxy clusters}",
      journal = {\mnras},
         year = "2015",
        month = "Nov",
       volume = {453},
       number = {3},
        pages = {3107-3119},
          doi = {10.1093/mnras/stv1852},
archivePrefix = {arXiv},
       eprint = {1508.05308},
 primaryClass = {astro-ph.CO},
       adsurl = {https://ui.adsabs.harvard.edu/abs/2015MNRAS.453.3107K}
}

@ARTICLE{OguriTakada11,
       author = {{Oguri}, Masamune and {Takada}, Masahiro},
        title = "{Combining cluster observables and stacked weak lensing to probe dark energy: Self-calibration of systematic uncertainties}",
      journal = {\prd},
         year = "2011",
        month = "Jan",
       volume = {83},
       number = {2},
          eid = {023008},
        pages = {023008},
          doi = {10.1103/PhysRevD.83.023008},
archivePrefix = {arXiv},
       eprint = {1010.0744},
 primaryClass = {astro-ph.CO},
       adsurl = {https://ui.adsabs.harvard.edu/abs/2011PhRvD..83b3008O}
}

@ARTICLE{Schaller15,
   author = {{Schaller}, M. and {Frenk}, C.~S. and {Bower}, R.~G. and {Theuns}, T. and 
	{Trayford}, J. and {Crain}, R.~A. and {Furlong}, M. and {Schaye}, J. and 
	{Dalla Vecchia}, C. and {McCarthy}, I.~G.},
    title = "{The effect of baryons on the inner density profiles of rich clusters}",
  journal = {\mnras},
archivePrefix = "arXiv",
   eprint = {1409.8297},
     year = 2015,
    month = sep,
   volume = 452,
    pages = {343-355},
      doi = {10.1093/mnras/stv1341},
   adsurl = {https://ui.adsabs.harvard.edu/abs/2015MNRAS.452..343S}
}

@ARTICLE{Murata18,
   author = {{Murata}, R. and {Nishimichi}, T. and {Takada}, M. and {Miyatake}, H. and 
	{Shirasaki}, M. and {More}, S. and {Takahashi}, R. and {Osato}, K.
	},
    title = "{Constraints on the Mass-Richness Relation from the Abundance and Weak Lensing of SDSS Clusters}",
  journal = {\apj},
archivePrefix = "arXiv",
   eprint = {1707.01907},
     year = 2018,
    month = feb,
   volume = 854,
      eid = {120},
    pages = {120},
      doi = {10.3847/1538-4357/aaaab8},
   adsurl = {http://adsabs.harvard.edu/abs/2018ApJ...854..120M}
}

@ARTICLE{Shirasaki16,
       author = {{Shirasaki}, Masato and {Nagai}, Daisuke and {Lau}, Erwin T.},
        title = "{Covariance in the thermal SZ-weak lensing mass scaling relation of galaxy clusters}",
      journal = {\mnras},
         year = 2016,
        month = aug,
       volume = {460},
       number = {4},
        pages = {3913-3924},
          doi = {10.1093/mnras/stw1263},
archivePrefix = {arXiv},
       eprint = {1603.08609},
 primaryClass = {astro-ph.CO},
       adsurl = {https://ui.adsabs.harvard.edu/abs/2016MNRAS.460.3913S}
}

@ARTICLE{Costanzi19SDSS,
       author = {{Costanzi}, M. and {Rozo}, E. and {Simet}, M. and {Zhang}, Y. and
         {Evrard}, A.~E. and {Mantz}, A. and {Rykoff}, E.~S. and {Jeltema}, T. and
         {Gruen}, D. and {Allen}, S. and {McClintock}, T. and {Romer}, A.~K. and
         {von der Linden}, A. and {Farahi}, A. and {DeRose}, J. and
         {Varga}, T.~N. and {Weller}, J. and {Giles}, P. and {Hollowood}, D.~L. and
         {Bhargava}, S. and {Bermeo-Hernandez}, A. and {Chen}, X. and
         {Abbott}, T.~M.~C. and {Abdalla}, F.~B. and {Avila}, S. and
         {Bechtol}, K. and {Brooks}, D. and {Buckley-Geer}, E. and
         {Burke}, D.~L. and others},
        title = "{Methods for cluster cosmology and application to the SDSS in preparation for DES Year 1 release}",
      journal = {\mnras},
         year = 2019,
        month = oct,
       volume = {488},
       number = {4},
        pages = {4779-4800},
          doi = {10.1093/mnras/stz1949},
archivePrefix = {arXiv},
       eprint = {1810.09456},
 primaryClass = {astro-ph.CO},
       adsurl = {https://ui.adsabs.harvard.edu/abs/2019MNRAS.488.4779C}
}

@misc{Corrfunc,
   author = {{Sinha}, M. and {Garrison}, L.},
   title = "{Corrfunc: Blazing fast correlation functions on the CPU}",
   howpublished = {Astrophysics Source Code Library},
   year = 2017,
   month = mar,
   archivePrefix = "ascl",
   eprint = {1703.003},
   adsurl = {http://adsabs.harvard.edu/abs/2017ascl.soft03003S}
}

@article{Weinberg13,
	Adsurl = {http://adsabs.harvard.edu/abs/2013PhR...530...87W},
	Archiveprefix = {arXiv},
	Author = {{Weinberg}, D.~H. and {Mortonson}, M.~J. and {Eisenstein}, D.~J. and {Hirata}, C. and {Riess}, A.~G. and {Rozo}, E.},
	Doi = {10.1016/j.physrep.2013.05.001},
	Eprint = {1201.2434},
	Journal = {\physrep},
	Month = sep,
	Pages = {87-255},
	Title = {{Observational probes of cosmic acceleration}},
	Volume = 530,
	Year = 2013}

@article{emcee,
	Adsurl = {http://adsabs.harvard.edu/abs/2013PASP..125..306F},
	Archiveprefix = {arXiv},
	Author = {{Foreman-Mackey}, D. and {Hogg}, D.~W. and {Lang}, D. and {Goodman}, J.},
	Doi = {10.1086/670067},
	Eprint = {1202.3665},
	Journal = {\pasp},
	Month = mar,
	Pages = {306-312},
	Primaryclass = {astro-ph.IM},
	Title = {{emcee: The MCMC Hammer}},
	Volume = 125,
	Year = 2013}

@article{Bhattacharya13,
	Adsurl = {http://adsabs.harvard.edu/abs/2013ApJ...766...32B},
	Archiveprefix = {arXiv},
	Author = {{Bhattacharya}, S. and {Habib}, S. and {Heitmann}, K. and {Vikhlinin}, A.},
	Doi = {10.1088/0004-637X/766/1/32},
	Eid = {32},
	Eprint = {1112.5479},
	Journal = {\apj},
	Month = mar,
	Pages = {32},
	Primaryclass = {astro-ph.CO},
	Title = {{Dark Matter Halo Profiles of Massive Clusters: Theory versus Observations}},
	Volume = 766,
	Year = 2013}

@article{Bahe12,
	Adsurl = {http://adsabs.harvard.edu/abs/2012MNRAS.421.1073B},
	Archiveprefix = {arXiv},
	Author = {{Bah{\'e}}, Y.~M. and {McCarthy}, I.~G. and {King}, L.~J.},
	Doi = {10.1111/j.1365-2966.2011.20364.x},
	Eprint = {1106.2046},
	Journal = {\mnras},
	Month = apr,
	Pages = {1073-1088},
	Primaryclass = {astro-ph.CO},
	Title = {{Mock weak lensing analysis of simulated galaxy clusters: bias and scatter in mass and concentration}},
	Volume = 421,
	Year = 2012}

@article{Voit05,
	Adsurl = {http://adsabs.harvard.edu/abs/2005RvMP...77..207V},
	Author = {{Voit}, G.~M.},
	Doi = {10.1103/RevModPhys.77.207},
	Eprint = {arXiv:astro-ph/0410173},
	Journal = {Reviews of Modern Physics},
	Month = apr,
	Pages = {207-258},
	Title = {{Tracing cosmic evolution with clusters of galaxies}},
	Volume = 77,
	Year = 2005}

@ARTICLE{Lesci22,
       author = {{Lesci}, G.~F. and {Marulli}, F. and {Moscardini}, L. and {Sereno}, M. and {Veropalumbo}, A. and {Maturi}, M. and {Giocoli}, C. and {Radovich}, M. and {Bellagamba}, F. and {Roncarelli}, M. and {Bardelli}, S. and {Contarini}, S. and {Covone}, G. and {Ingoglia}, L. and {Nanni}, L. and {Puddu}, E.},
        title = "{AMICO galaxy clusters in KiDS-DR3: Cosmological constraints from counts and stacked weak lensing}",
      journal = {\aap},
         year = 2022,
        month = mar,
       volume = {659},
          eid = {A88},
        pages = {A88},
          doi = {10.1051/0004-6361/202040194},
archivePrefix = {arXiv},
       eprint = {2012.12273},
 primaryClass = {astro-ph.CO},
       adsurl = {https://ui.adsabs.harvard.edu/abs/2022A&A...659A..88L}
}

@preamble{ "\newcommand{\apjl}{Astrophys. J. Lett.}" }

@preamble{ "\newcommand{\apjs}{Astrophys. J. Suppl.}" }

@preamble{ "\newcommand{\mnras}{Mon. Not. R. Astron. Soc.}" }

@preamble{ "\newcommand{\jcap}{J. Cosmol. Astropart. Phys.}" }

@preamble{ "\newcommand{\aap}{Astron. Astrophys.}" }

@preamble{ "\newcommand{\rmxaa}{Revista Mexicana de Astronom{\'i}a y Astrof{\'i}sica}" }

@preamble{ "\newcommand{\phiv}{$\phi$}" }

@preamble{ "\newcommand{\apss}{Astrophys. Space Sci.}" }

@preamble{ "\newcommand{\pasp}{PASP}" }

@preamble{ "\newcommand{\nar}{New Astronomy Reviews}" }

@preamble{ "\newcommand{\pasj}{PASJ}" }

@preamble{ "\newcommand{\araa}{ARAA}" }

@preamble{ "\newcommand{\aapr}{AAPR}" }

@preamble{ "\newcommand{\aj}{AJ}" }

@preamble{ "\newcommand{\physrep}{Phys. Rep.}" }

@preamble{ "\newcommand{\ssr}{Space Sci. Rev.}" }
\end{document}